\documentclass[showpacs,preprintnumbers,amsmath,amssymb,prb,twocolumn]{revtex4-1}

\usepackage{txfonts}
\usepackage{epsfig}
\begin{document}

\title{Josephson Effects in Three-Band Superconductors with Broken Time-Reversal Symmetry}

\author{Zhao Huang\(^{1,2}\) and Xiao Hu\(^{1,2}\)}
 \email{hu.xiao@nims.go.jp.}

\affiliation{\(^{1}\) International Center for Materials Nanoarchitectonics (WPI-MANA), National Institute for Materials Science, Tsukuba 305-0044, Japan\\
\(^{2}\)Graduate School of Pure and Applied Sciences, University of Tsukuba, Tsukuba 305-8571, Japan}

\begin{abstract}
In superconductors with three or more bands, time-reversal symmetry (TRS) may be broken in
the presence of repulsive interband couplings, resulting in a pair of degenerate states
characterized by opposite chiralities. We consider a Josephson junction between a three-band
superconductor with broken TRS and a single-band superconductor. Phenomena such as asymmetric
critical currents, subharmonic Shapiro steps and symmetric Fraunhhofer patterns are revealed
theoretically. Existing experimental results are discussed in terms of the present work.
\end{abstract}

\date{\today}

\maketitle

\maketitle
The Josephson effect is a remarkable macroscopic tunneling phenomenon associated with broken gauge symmetry in a superconducting state \cite {Josephson62}. When phase difference $\Delta\varphi$ exists between two superconductors connected by a weak link, dc supercurrent flows through the junction with zero voltage bias. The detailed form of current phase relation (CPR) depends on the materials and geometries of the weak links, while the ac Josephson relation is given by  $\partial_t(\Delta\varphi)=2eV/\hbar$.

Because the Josephson effect is due to interference between wave functions of two superconductors that are weakly linked, it carries the information of gap structures. Therefore, it is widely used as a tool to detect the pairing symmetry in an unconventional superconductor. For example, the half-flux quanta observed in the tricrystal junction in high-temperature cuprate superconductor serves as the best evidence for the d-wave pairing symmetry \cite {Tsuei00}.

There has been a rapidly growing interest in multi-band superconductors stimulated by discovery of $\textrm{MgB}_2$ and iron pnictides \cite {Nagamatsu01,Kamihara08,Kuroki08}, in which superconductivity in different bands couples through interband couplings. Let us first see a two-band superconductor with interband Cooper-pair scattering as in previous works \cite{Suhl59,Kondo63,Leggett66,Gurevich07}. An attractive coupling leads to two parallel order parameters while a repulsive coupling gives opposite sign of two order parameters. The situation becomes much different when there are three bands, where a frustrated state can emerge as a compromise of three repulsive interband couplings. In this case, interband phase differences are neither 0 nor $\pi$, leading to time-reversal symmetry (TRS) breaking \cite {Agterberg99, Stanev10, Tanaka10, Carlstrom11, Yanagisawa12, Dias11, Hu12, Lin12, Maiti13}. The time-reversal symmetry broken (TRSB) state can be realized in a multi-band superconductor even with all the gap functions of s-wave symmetry, which distinguishes it from magnetic superconductors and chiral p-wave superconductors.

\begin{figure}[t]
\psfig{figure=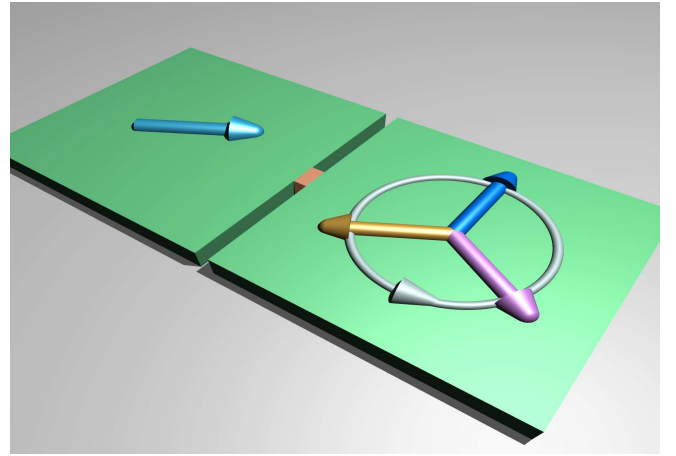,width=\columnwidth} \caption{\label{fig1} (color online) Schematics of point contact junction between single-band superconductor and three-band TRSB superconductor. Arrows indicate phases of gap functions and white circle with rotating direction represents the chirality.}
\end{figure}

A hopeful candidate of this TRSB state is the iron-based superconductor with at most five gaps originating from the five Fe 3d orbitals \cite {Kontani10}. In $\textrm{Ba}_{0.6}\textrm{K}_{0.4}\textrm{Fe}_2\textrm{As}_2$, angle resolved photoemission spectroscopy (ARPES) measurements observed four different gaps at two electron-like and two hole-like Fermi pockets \cite {Nakayama09}. Sign reversals between Cooper pairing of electron pockets and hole pockets caused by spin fluctuations were discussed \cite {Mazin08,Paglione10}. A sign reversal between two strong hole pockets has also been suggested in $\textrm{KFe}_2\textrm{As}_2$ \cite {Okazaki12,Maiti13,Maiti12}. Therefore, it is of interest to investigate the consequences of repulsive inter-band couplings in a general point of view.

It has been suggested that spontaneous broken TRS is accompanied by novel phenomena such as fractional vortices \cite{Sigrist95,Garaud11}, modified phase slip \cite{Sigrist98}, spontaneous supercurrent and self-induced flux\cite{Tsuei00,Ng09}, massless Leggett mode \cite{Lin12,Stanev12}, mixed phase density mode \cite{Silaev13} and vortex clusters \cite{Takahashi14}. In this paper, we investigate the tunneling phenomena in a Josephson junction connecting a single-band superconductor and a three-band TRSB superconductor as shown in Fig. \ref{fig1}.

We consider a point contact junction \cite{Kulik70} with the size much smaller than the BCS coherence length. We model the SNS junction by a step change in the gap functions
\begin{equation}
\Delta(x)=\left\{
\begin{array}{lll}
\Delta_0, &x<0;\\
0, &x=0;\\
\{\Delta_1,\Delta_2,\Delta_3\}, &x>0,
\end{array}
\right.
\end{equation}
with the origin at the point contact in Fig. \ref{fig1}; for simplicity all gaps are taken as s-wave pairing. In the bulk on the right hand side, $\Delta_1,\Delta_2$ and $\Delta_3$ are affected by the interband Cooper-pair scattering processes, and nontrivial phase differences exist among $\Delta_1,\Delta_2$ and $\Delta_3$\cite{Hu12}. By extending previous works for two-band case \cite{Ng09,Sperstad09}, the full Bogoliubov-de Gennes (BdG) equations are given by
\begin{equation}\label{hi}
\left( \begin{array}{lll}
H_1+T_{11} &T_{12} &T_{13}\\
T_{12} &H_2+T_{22} &T_{23}\\
T_{13} &T_{23} &H_3+T_{33}
\end{array} \right)
\left( \begin{array}{lll}
\psi_1\\
\psi_2\\
\psi_3
\end{array} \right)
=E
\left( \begin{array}{lll}
\psi_1\\
\psi_2\\
\psi_3
\end{array} \right),
\end{equation}
with $T_{jk}=t_{jk}\delta(x)\sigma_z$, $\psi_j=(u_{j\uparrow}(x),v_{j\downarrow}(x))^T$ and
\begin{equation}\label{hj}
H_j=
\left( \begin{array}{ll}
-{\hbar^2\partial^2_x}/{2m}-E_F &\Delta_j\Theta(x)+\Delta_0\Theta(-x)\\
\Delta_j^*\Theta(x)+\Delta_0^*\Theta(-x) &{\hbar^2\partial^2_x}/{2m}+E_F
\end{array} \right)
\end{equation}
with $\sigma_z$ the Pauli matrix, $\Theta(x)$ the Heaviside function and $E_F=\hbar^2k_F^2/2m$. Here $\psi_j$ are written in the Nambu spinor notation and $T_{jk}$ are the intraband $(j=k)$ and interband $(j\neq k)$ single-particle scattering terms. The BdG equations describe the transporting process of Copper pairs through the Andreev reflections \cite{Blonder82} in the ballistic limit.

In order to calculate the Andreev levels, we consider the right moving quasiparticles in the three-band superconductor on the right-hand side and the left moving quasiparticles in the single-band superconductor on the left-hand side \cite{Kulik70}. The solution has the form $\psi_j=\psi_{j-}\Theta(-x)+\psi_{j+}\Theta(x)$ with
\begin{equation}\label{excitations}
\begin{split}
\psi_{j-}=a_{j-}
\left( \begin{array}{ll}
u_0e^{i\varphi_0}\\
v_0
\end{array} \right)e^{-ik_Fx}
+b_{j-}
\left( \begin{array}{ll}
v_0e^{i\varphi_0}\\
u_0
\end{array} \right)e^{ik_Fx},\\
\psi_{j+}=a_{j+}
\left( \begin{array}{ll}
u_je^{i\varphi_j}\\
v_j
\end{array} \right)e^{ik_Fx}
+b_{j+}
\left( \begin{array}{ll}
v_je^{i\varphi_j}\\
u_j
\end{array} \right)e^{-ik_Fx},
\end{split}
\end{equation}
where
\begin{flalign}\label{uv}
u_j=\sqrt{\frac{1}{2}\left(1+\sqrt{1-\left(\frac{|\Delta_j|}{E}\right)^2}\right)},v_j=\sqrt{\frac{1}{2}\left(1-\sqrt{1-\left(\frac{|\Delta_j|}{E}\right)^2}\right)}
\end{flalign}
with $j=1,2$ and $3$.
To obtain Eq. (\ref{excitations}), we have assumed the Andreev approximation $E,\Delta_j\ll E_F$, and thus the wave vectors simply read $\pm k_F$. While Eq. (\ref{excitations}) shares the same form with the two-band superconductors \cite{Sperstad09}, the physics is quite different as will be revealed below due to the broken TRS.

From Eq. (\ref{excitations}), the boundary conditions are given by
\begin{equation}\label{boundary}
\begin{split}
&\psi_{j-}(0)=\psi_{j+}(0),\\
\left.\left(\frac{\partial\psi_{j+}}{\partial x}-\frac{\partial\psi_{j-}}{\partial x}\right)\right|_{x=0}&=\frac{2m}{\hbar^2}
\left[t_{jj}\psi_{j-}(0)+t_{jk}\psi_{k-}(0)+t_{jl}\psi_{l-}(0)\right].
\end{split}
\end{equation}

Here we consider a simple case with $|\Delta_0|=|\Delta_j|=|\Delta|$ in the ballistic limit $(t_{jk}=0)$. Results for a general case are available as a supplementary online material \cite{supplementary}. From Eq. (\ref{boundary}), we obtain the Andreev spectra as
\begin{equation}
E_j^{\pm}=\pm\left|\Delta\right|\cos(\varphi_{j0}/2)
\end{equation}
with $\varphi_{j0}=\varphi_j-\varphi_0$. Supercurrent in each channel is given by \cite{Beenakker91}
\begin{equation}\label{CPRformula}
I_{j}=\frac{2e}{\hbar}\frac{\partial E_j^{+}}{\partial \varphi_{j0}}f(E_j^{+})+\frac{2e}{\hbar}\frac{\partial E_j^{-}}{\partial \varphi_{j0}}f(E_j^{-})
\end{equation}
where $f(E_j)$ is the Fermi-Dirac distribution function. Thus we obtain the total current
\begin{equation}
\begin{split}\label{pcurrent}
I_s(\varphi)&=\frac{e|\Delta|}{\hbar}\sin({\varphi}/{2})\tanh\frac{|\Delta|\cos({\varphi}/{2})}{2k_BT}\\
&+\frac{e|\Delta|}{\hbar}\sin[{(\varphi+\varphi_{21})}/{2}]\tanh\frac{|\Delta|\cos[(\varphi+\varphi_{21})/{2}]}{2k_BT}\\
&+\frac{e|\Delta|}{\hbar}\sin[{(\varphi+\varphi_{31})}/{2}]\tanh\frac{|\Delta|\cos[(\varphi+\varphi_{31})/{2}]}{2k_BT},
\end{split}
\end{equation}
where $\varphi=\varphi_{10}$, and $\varphi_{21}$ and $\varphi_{31}$ are interband phase differences for the three-band superconductor.

\begin{figure}[t]
\psfig{figure=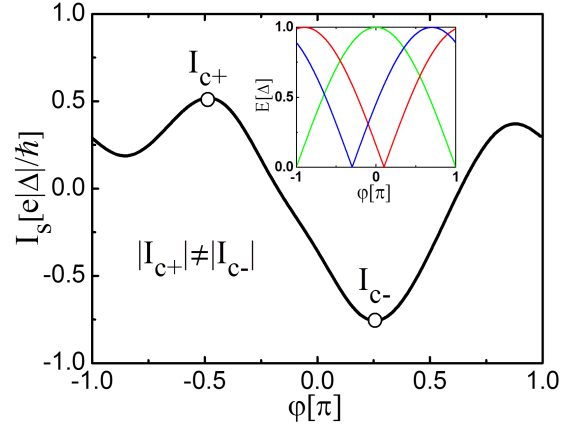,width=8.0cm} \caption{\label{fig2} (color online) Current phase relation of ballistic point-contact junction between single-band and three-band superconductor for parameters $\varphi_{21}=0.9\pi$, $\varphi_{31}=1.3\pi$ and $ T=0.2\Delta/k_B$. Inset: Andreev spectra for the same parameter set.}
\end{figure}

In a three-band superconductor with interband attractions, all gap functions take the same phase which reserves the TRS. However, when the interband couplings are repulsive, gap functions have to take different phases with nontrivial phase differences, which breaks the TRS \cite{Agterberg99, Stanev10, Tanaka10, Carlstrom11, Yanagisawa12, Dias11, Hu12, Lin12, Maiti13}.

To be specific, we consider a case with equal gap amplitudes albeit different interband couplings, such that $\varphi_{21}=0.9\pi$ and $\varphi_{31}=1.3\pi$. The current phase relation (CPR) for this system is displayed in Fig. 2. It is interesting to find that critical currents in the two opposite directions are unequal.

The reason for the asymmetry in the critical currents can be found from the symmetry of CPR. In a conventional time-reversal symmetry reserved (TRSR) superconductor (three-band: $\varphi_{jk}=0$ or $\pi$), the same critical currents in two directions is a direct consequence of anti-reversal symmetry of CPR $I(-\varphi)=-I(\varphi)$ as can be seen in Eq. (\ref{pcurrent}). This property is protected by TRS \cite{Golubov04}. However in a TRSB state under concern, this anti-reversal symmetry is broken as in Eq. (\ref{pcurrent}) since $\varphi_{21},\varphi_{31}\neq0,\pi$. The asymmetric critical currents can also be understood from the absence of symmetry in the Andreev spectra as shown in Fig. \ref{fig2}.

The relation between the broken TRS and asymmetric critical currents can be seen in another way. Upon the time-reversal transformation to the whole system, supercurrent reverses its direction and the superconducting state changes according to $(\varphi_j\rightarrow-\varphi_j)$. In a TRSB state, the superconducting state upon TRS operation results in another state with chirality opposite to the original one. Therefore the same amount of supercurrents can only be guaranteed by two different states, which are not connected to each other. It is clearly not the case in the TRSR state.

As temperature approaches $T_c$, the gap function is suppressed and Eq. (\ref{pcurrent}) is approximately reduced to
\begin{equation}\label{sincurrent}
I_s(\varphi)=\frac{e|\Delta|^2}{4\hbar k_B T}[\sin\varphi+\sin(\varphi+\varphi_{21})+\sin(\varphi+\varphi_{31})].
\end{equation}
A translational anti-symmetry $I_s(\varphi)=-I_s(\varphi+\pi)$ appears, which makes the two critical currents in the two opposite directions equal to each other. At low temperatures, the translational antisymmetry is destroyed by high harmonics additional to those in Eq. (\ref{sincurrent}), which realize the asymmetric critical currents discussed above.

Now we analyze the response of the TRSB state to a microwave irradiation. For this purpose, we rewrite the CPR Eq. (\ref{pcurrent}). In the three tunneling channels, we have $I_j(-\varphi_{j0})=-I_j(\varphi_{j0})$ and $I_j(\varphi_{j0}+2\pi)=I_j(\varphi_{j0})$ according to Eq. (\ref{CPRformula}). Therefore, $I_j$ can be expanded into a Fourier series with only sine functions, and thus the CPR can be written as
\begin{equation}\label{3CFourier}
\begin{split}
I_s(\varphi)=\sum_{n=1}^{+\infty}[I_{1n}\sin n\varphi+I_{2n}\sin n(\varphi+\varphi_{21})+I_{3n}\sin n(\varphi+\varphi_{31})],
\end{split}
\end{equation}
where
\begin{equation}
I_{jn}=\frac{1}{\pi}\int_{-\pi}^{\pi}\frac{e|\Delta|}{\hbar}\sin(\xi/2)\tanh\frac{|\Delta|\cos(\xi/2)}{2k_BT}\sin n\xi d\xi
\end{equation}
with $j=1,2$ and $3$.
It is noticed that the supercurrents carried by the three channels may enhance and suppress each other, depending on the phase differences and the order of Fourier components. To illustrate this effect clearly, we consider an isotropic state with $|\Delta_1|=|\Delta_2|=|\Delta_3|$ and $\varphi_{21}=\varphi_{32}=2\pi/3$ at the right-hand side of the junction. It is interesting to observe that a complete cancelation takes place for $3n+1$ and $3n+2$ for n=0,1,2, ... , and the CPR is reduced to
\begin{equation}
I_s(\varphi)=I_c\sin3\varphi.
\end{equation}

\begin{figure}[t]
\psfig{figure=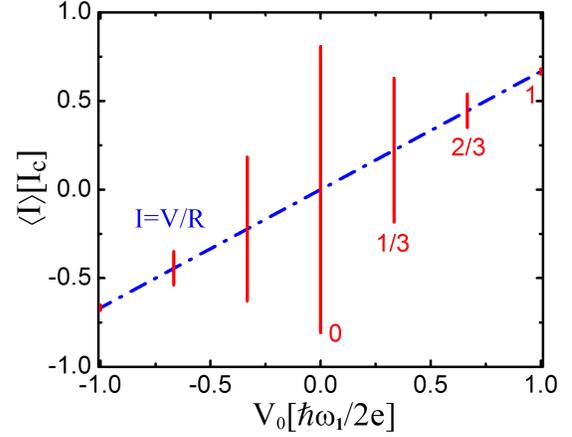,width=8.0cm} \caption{\label{fig3} (color online) Subharmonic Shapiro steps of overdamped junction between single-band and three-band isotropic TRSB superconductor ($|\Delta_1|=|\Delta_2|=|\Delta_3|$ and $\varphi_{21}=\varphi_{32}=2\pi/3$) driven by ac voltage source. The parameters are $V_{1}=0.15\hbar\omega_1/e$ and $R=0.75\hbar\omega_1/eI_c$.}
\end{figure}

Now we take the applied voltage with the form $V=V_0+V_1\cos\omega_1 t$. The Josephson current is given by
\begin{equation}
\begin{split}
I_s(\varphi)=I_c\sum_{m=-\infty}^{+\infty}(-1)^mJ_m(\frac{6eV_1}{\hbar\omega_1})\sin(\delta_0+3\omega_0t-m\omega_1t)
\end{split}
\end{equation}
with $J_m$ the Bessel function of the first kind, $\delta_0$ an arbitrary phase and $\omega_0=2eV_0/\hbar$. It is easy to find that the supercurrent contains a dc component when
\begin{equation}
V_0=\frac{m}{3}\frac{\hbar\omega_1}{2e}.
\end{equation}
The fractional numbers ${m}/{3}$ indicate the subharmonic Shapiro steps.

We assume that junction is overdamped with a shunted resistance $R$, and thus the total dc current on the $m$th subharmonic Shapiro step can take any value in the range
\begin{equation}
\frac{V_0}{R}-I_cJ_m(\frac{6eV_1}{\hbar\omega_1})\leq \langle I\rangle \leq \frac{V_0}{R}+I_cJ_m(\frac{6eV_1}{\hbar\omega_1})
\end{equation}
as shown in Fig. \ref{fig3}.

For general TRSB states where cancelations are not complete, $1/2$ Shapiro step appears. However, because the $1/2$ step is still suppressed and the $1/3$ step is enhanced, we expect that the $1/3$ step is comparable or even larger than the $1/2$ step.

Now let us see the Fraunhofer interference in an extended junction between a single-band superconductor and a three-band TRSB superconductor. We adopt the general expression for Josephson currents \cite{Golubov04,Tanaka94} $I_s(\varphi)$ in Eq. (\ref{3CFourier}). In the presence of a flux $\Phi$ penetrating into the junction, the Josephson current is given by
\begin{equation}\label{Fraunhofer}
\begin{split}
&I_{\Phi}(\varphi)=\frac{1}{L}\int_0^L I_s(\varphi+\frac{2\pi\Phi}{\Phi_0}\frac{x}{L})dx=\sum_n\frac{\sin({n\pi\Phi}/{\Phi_0})}{{n\pi\Phi}/{\Phi_0}}\\
&\times[I_{1n}\sin n\varphi+
I_{2n}\sin n(\varphi+\varphi_{21})+
I_{3n}\sin n(\varphi+\varphi_{31})].
\end{split}
\end{equation}
We find that the supercurrent is symmetric with respect to the direction of magnetic flux, which indicates that TRSB states do not couple with magnetic fields.

Although we have focused on point-contact junction in the above investigations, it is clear that the main results remain valid for
junctions with extensions since the most essential ingredient is the existence of TRSB state characterized by
nontrivial phase differences among condensates, in which the high harmonics in Josephson current become more
prominent.

There is a special case where component-2 and component-3 are equivalent, which results in $\varphi_{21}=-\varphi_{31}$, and the critical currents in opposite directions are equal according to Eq.~(\ref{pcurrent}), albeit the broken TRS. We notice that the case  $\varphi_{21}=-\varphi_{31}$ generated by two equivalent components is accidental without symmetry protection. When temperature changes, gap amplitudes and/or interband couplings become different for the two components, for which asymmetric critical currents appear. In the case that component-2 coincides with component-3, we may distinguish the TRSB state from TRSR state by using the Shapiro steps. For TRSB state, 1/3-step is larger than or comparable to 1/2-step (Fig. 3 is an extreme case where 1/2-step is suppressed to zero since the three components are equivalent), while in the TRSR state 1/2-step is expected larger than 1/3-step.

It is intriguing to notice that asymmetric critical currents have been observed in a hybrid junction between a single-band superconductor  $\textrm{PbIn}$ and an iron-based superconductor $\textrm{BaFe}_{2-x}\textrm{Co}_{x}\textrm{As}_2$ \cite{Schmidt10,Doring12}. The difference between two critical currents is well beyond the experimental precision. Subharmonic Shapiro steps were also detected in the same setup, indicating the importance of high-order harmonics in the Josephson current. While the asymmetric critical currents was explained by presuming vortices accidentally trapped in one of the two superconductors \cite{Berdiyorov11}, we wish to point out the two phenomena observed in the experiments can be understood in terms of the existence of a TRSB state as discussed in the present work. Further careful investigations are highly anticipated to make the situation clear. For example, the Fraunhofer pattern is to be measured, which should be asymmetric with respect to the flux direction if some vortices are trapped in one of the two superconductors \cite{Golod10}, while the TRSB state corresponds to a symmetric one.

We notice that there is another similar experiment for a junction between $\textrm{Pb}$ and an iron pnictide $\textrm{Ba}_{1-x}\textrm{K}_x\textrm{Fe}_2\textrm{As}_2$ with $x=0.29$ and $x=0.49$, where symmetric critical currents have been observed \cite{Zhang09}. It is possible that in order to see the TRSB state, and thus the asymmetric critical currents, the hole doping rate is to be tuned to the overdoping regime, according to a recent theoretical work \cite{Maiti13}.

To summarize, in superconductors with three or more bands, time-reversal symmetry may be broken in the presence of repulsive interband couplings, characterized by
nontrivial phase differences among condensates. Due to the broken time-reversal symmetry, asymmetric critical currents appear in a Josephson junction between
a single-band and a multi-band superconductor. Subharmonic Shapiro steps become more prominent since the tunneling currents carried by different bands may
cancel each other, which reduces the sizes of integer Shapiro steps.
\vspace{2mm}

\leftline{{\textbf{Acknowledgements}}}

\vspace{2mm}
The authors are grateful for T. Kawakami, Q. F. Liang, Y. Takahashi and Z. Wang for discussions. This work was supported by the WPI initiative on Materials Nanoarchitectonics, and partially by the Grant-in-Aid for Scientific Research (No. 25400385), the Ministry of Education, Culture, Sports, Science and Technology of Japan.


%

\begin{widetext}

\title{\LARGE{Supplementary information to the manuscript: Josephson Effects in Three-Band Superconductors with Broken Time-Reversal Symmetry}}

\vspace{10mm}

\centerline{{\LARGE{Supplementary information to the manuscript: Josephson Effects in}}}
\centerline{{\LARGE{Three-Band Superconductors with Broken}}}
\centerline{{\LARGE{Time-Reversal Symmetry}}}

\vspace{-1mm}
\vspace{5mm}
\centerline{\large{\textbf{Tunneling current when gap amplitudes are different}}}
\vspace{5mm}

In the main context, we have considered the case of $|\Delta_0|=|\Delta_j|$. In this supplementary, we consider the general situation with $|\Delta_0|\neq |\Delta_j|$, in which Josephson current is contributed from not only discrete Andreev levels below the two gaps but also intermediate continuum levels between the two gaps with a formalism similar to the single-band system \cite{Chang94}.

We first calculate the supercurrent carried by the Andreev levels. As in the main text, the wave functions at two sides of the junction and boundary conditions are given by
\begin{equation}\label{excitations1}
\begin{split}
\psi_{j-}=a_{j-}
\left(\begin{array}{ll}
u_0e^{i\varphi_0}\\
v_0
\end{array}\right)e^{-ik_Fx}
+b_{j-}
\left(\begin{array}{ll}
v_0e^{i\varphi_0}\\
u_0
\end{array}\right)e^{ik_Fx},\\
\psi_{j+}=a_{j+}
\left(\begin{array}{ll}
u_je^{i\varphi_j}\\
v_j
\end{array}\right)e^{ik_Fx}
+b_{j+}
\left(\begin{array}{ll}
v_je^{i\varphi_j}\\
u_j
\end{array}\right)e^{-ik_Fx},
\end{split}
\end{equation}
and
\begin{equation}\label{boundary1}
\begin{split}
&\psi_{j-}(0)=\psi_{j+}(0),\\
\left.\left(\frac{\partial\psi_{j+}}{\partial x}-\frac{\partial\psi_{j-}}{\partial x}\right)\right|_{x=0}&=\frac{2m}{\hbar^2}
\left[t_{jj}\psi_{j-}(0)+t_{jk}\psi_{k-}(0)+t_{jl}\psi_{l-}(0)\right],
\end{split}
\end{equation}
which lead to
\begin{equation}\label{Andreev1}
\left(\begin{array}{lll}
M_1 &C_{12} &C_{13}\\
C_{12} &M_2 &C_{23}\\
C_{13} &C_{23} &M_{3}
\end{array}\right)
\left(\begin{array}{lll}
D_1\\
D_2\\
D_3
\end{array}\right)
=0,
\end{equation}
with $D_j=(a_{j+},b_{j+},a_{j-},b_{j-})^T$,
\begin{equation*}
M_j=
\left(\begin{array}{llll}
u_je^{i\varphi_{j0}} &v_je^{i\varphi_{j0}} &-u_0 &-v_0\\
v_j &u_j &-v_0 &-u_0\\
u_je^{i\varphi_{j0}} &-v_je^{i\varphi_{j0}} &(1-T_{jj})u_0 &-(1+T_{jj})v_0\\
v_j &-u_j &(1-T_{jj})v_0 &-(1+T_{jj})u_0
\end{array}\right),
\end{equation*}
and
\begin{equation*}
C_{jk}=
\left(\begin{array}{llll}
0 &0 &0 &0\\
0 &0 &0 &0\\
0 &0 &-T_{jk}u_0 &-T_{jk}v_0\\
0 &0 &-T_{jk}v_0 &-T_{jk}u_0
\end{array}\right)
\end{equation*}
with $T_{jk}={2mt_{jk}}/{i\hbar^2 k_F}$.

Non-zero solutions are available when the determinant of the square matrix in Eq. (\ref{Andreev1}) is zero, which results in the Andreev spectra $E_j^{\pm}(\varphi_{j0})$. The tunneling current can be calculated with the formula
\begin{equation}\label{CPRformula1}
I_{A}=\sum_{j=1,2,3}\frac{2e}{\hbar}\frac{\partial E_j^{+}}{\partial \varphi_{j0}}f_{E_j^{+}}+\frac{2e}{\hbar}\frac{\partial E_j^{-}}{\partial \varphi_{j0}}f_{E_j^{-}},
\end{equation}
where $f_{E_j^+}$ and $f_{E_j^-}$ are the Fermi-Dirac distribution function.

The analytical expression of $I_A$ is available for a ballistic junction with $t_{jk}=0$, where the square matrix in Eq. (\ref{Andreev1}) is reduced to three $4\times4$ diagonal block matrices. The Andreev spectra are obtained as
\begin{equation}
E^{\pm}_j=\pm\frac{|\Delta_0||\Delta_j|\sin\varphi_{j0}}{\sqrt{|\Delta_0|^2+|\Delta_j|^2-2|\Delta_0||\Delta_j|\cos\varphi_{j0}}}
\end{equation}
with the restriction $\cos\varphi_{j0}\leq |\Delta_0/\Delta_j|$, which carry the tunneling current
\begin{equation}
\begin{split}
I_A=\sum_{j=1,2,3}\left[\frac{2e}{\hbar}\frac{|\Delta_0|^2|\Delta_j|^2-\left(|\Delta_0|^2+|\Delta_j|^2-|\Delta_0||\Delta_j|\cos\varphi_{j0}\right)|\Delta_0||\Delta_j|\cos\varphi_{j0}}{\left(|\Delta_0|^2+|\Delta_j|^2-|\Delta_0||\Delta_j|\cos\varphi_{j0}\right)^{3/2}}\right.\\
\left.\times\tanh\frac{|\Delta_0||\Delta_j|\sin\varphi_{j0}}{2k_BT\sqrt{|\Delta_0|^2+|\Delta_j|^2-|\Delta_0||\Delta_j|\cos\varphi_{j0}}}\right].
\end{split}
\end{equation}

\begin{figure}[t]
\psfig{figure=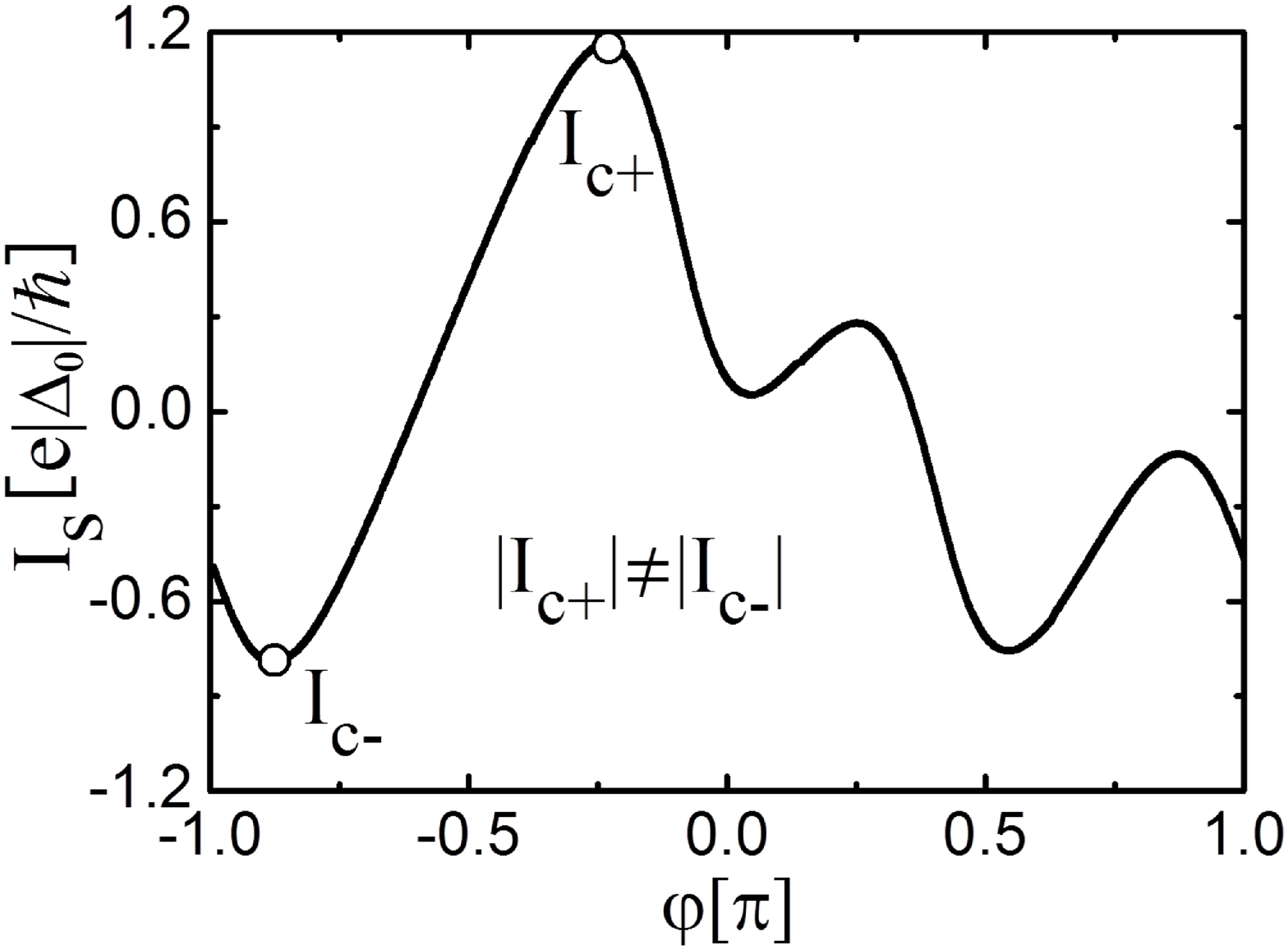,width=\columnwidth} \caption{\label{fig4} (color online) Current phase relation of point-contact junction between single-band and three-band TRSB superconductor with different gap amplitudes. Here we have $|\Delta_1|=1.2|\Delta_0|$, $|\Delta_2|=1.3|\Delta_0|$, $|\Delta_3|=1.5|\Delta_0|$, $\varphi_{21}=0.6\pi$, $\varphi_{31}=1.1\pi$, $t_{11}=t_{22}=t_{33}=0.2$ and $t_{12}=t_{13}=t_{23}=0.1$. All $t_{jk}$ are in unite of $2m/\hbar^2k_F$.}
\end{figure}

Now we calculate the tunneling current flowing in the energy region between the two gaps. To be specific, we take $|\Delta_0|<|\Delta_j|$. In this energy region, electron-like and hole-like quasiparticles are injected from the left single-band superconductor and transmitted into the right three-band superconductor, contributing to the tunneling current.

The wave function in the left superconductor is given by
\begin{equation}
\psi_{j-}=
\left(\begin{array}{ll}
u_0e^{i\varphi_0}\\
v_0
\end{array}\right)e^{ik_Fx}
+a_{j-}
\left(\begin{array}{ll}
u_0e^{i\varphi_0}\\
v_0
\end{array}\right)e^{-ik_Fx}
+b_{j-}
\left(\begin{array}{ll}
v_0e^{i\varphi_0}\\
u_0
\end{array}\right)e^{ik_Fx},
\end{equation}
where the first term is the incident electron-like quasiparticle which is absent in Eq. (\ref{excitations1}), the second term is the reflected electron-like quasiparticle and the third term is the reflected hole-like quasiparticle.
The transmitted wave $\psi_{j+}$ maintains the form
\begin{equation}
\psi_{j+}=a_{j+}
\left(\begin{array}{ll}
u_je^{i\varphi_j}\\
v_j
\end{array}\right)e^{ik_Fx}
+b_{j+}
\left(\begin{array}{ll}
v_je^{i\varphi_j}\\
u_j
\end{array}\right)e^{-ik_Fx}.
\end{equation}
By using the boundary conditions in Eq. (\ref{boundary1}), we obtain
\begin{equation}\label{intermediate}
\left(\begin{array}{lll}
M_1 &C_{12} &C_{13}\\
C_{12} &M_2 &C_{23}\\
C_{13} &C_{23} &M_{3}
\end{array}\right)
\left(\begin{array}{lll}
D_1\\
D_2\\
D_3
\end{array}\right)
=
\left(\begin{array}{lll}
Q_1\\
Q_2\\
Q_3
\end{array}\right),
\end{equation}
with $Q_j=\left(u_0,v_0,(1+T_{jj}+T_{jk}+T_{jl})u_0,(1+T_{jj}+T_{jk}+T_{jl})v_0\right)^T$, from where we can obtain $(a_{j+},b_{j+},a_{j-},b_{j-})$. It is noted that Eq. (\ref{CPRformula1}) is not convenient for calculating the tunneling current in the present continuum energy regime. Instead we integrate the electrical current density as
\begin{equation}\label{intercurrent}
I_I^e=\sum_{j=1,2,3}e\frac{\hbar k_F}{m}\left(\int_{|\Delta_0|}^{|\Delta_j|}+\int_{-|\Delta_j|}^{-|\Delta_0|}\right)\left(|a_{j+}|^2-|b_{j+}|^2\right){N_{s-}}f_{E_j}dE_j,
\end{equation}
where $N_{s-}$ is the density of states for the electron-like quasiparticles in the left superconductor and $f_{E_j}$ is the Fermi-Dirac distribution function. A similar calculation can be done for the incident hole-like quasiparticles, which contribute a current $I_I^h$. The total current carried by the intermediate energy region is thus given by
\begin{equation}\label{intertotal1}
I_I=I_I^e+I_I^h.
\end{equation}

The ballistic case can be solved analytically and the tunneling current is
\begin{equation}
\begin{split}
I_I=\sum_{j=1,2,3}\left[\frac{2e}{h}\int_{|\Delta_0|}^{|\Delta_j|}\sqrt{1-\left|\frac{\Delta_0}{E_j}\right|^2}\left(\frac{1}{1-\left|\frac{\Delta_0}{E_j}\right|\cos\left[\varphi_{j0}-\cos^{-1}\left|\frac{E_j}{\Delta_j}\right|\right]}\right.\right.\\
\left.\left.-\frac{1}{1-\left|\frac{\Delta_0}{E_j}\right|\cos\left[\varphi_{j0}+\cos^{-1}\left|\frac{E_j}{\Delta_j}\right|\right]}\right)\tanh\frac{E_j}{2k_BT}dE_j\right].
\end{split}
\end{equation}
The calculation result can be extended to the $|\Delta_0|>|\Delta_j|$, by simply reversing the limits of integration.

For a non-ballistic junction with scattering centers at the interface $(t_{jk}\neq0)$, numerical calculations are required and the current phase relation is shown in Fig. \ref{fig4}. An asymmetry in critical currents is found when the three-band superconductor stays at a TRSB state where $\varphi_{21},\varphi_{31}\neq0,\pi$.

\end{widetext}

\end{document}